%% file: JFM.tex
\documentclass[]{jfm}

\usepackage{graphicx}
\usepackage{newtxtext}
\usepackage{newtxmath}
\usepackage{natbib}
\usepackage{hyperref}
\hypersetup{
    colorlinks = true,
    urlcolor   = blue,
    citecolor  = black,
}

\newcommand{\RomanNumeralCaps}[1]
\linenumbers

\newcommand{\pardt}[1]{\frac{\partial #1}{\partial t}}
\newcommand{\pardT}[1]{\frac{\partial #1}{\partial T}}

\newcommand{\pardx}[1]{\frac{\partial #1}{\partial x}}

\newcommand{\pardr}[1]{\frac{\partial #1}{\partial r}}

\newcommand{\pardtheta}[1]{\frac{\partial #1}{\partial \theta}}
\newcommand{\pardphi}[1]{\frac{\partial #1}{\partial \phi}}

\newcommand{\lapp}{\nabla^2_\mathbf{p}}

\newcommand{\delp}{\bnabla_\mathbf{p}}

\newcommand{\pvec}{\mathbf{p}}
\newcommand{\xvec}{\mathbf{x}}
\newcommand{\uvec}{\mathbf{u}}

\newcommand\Pe{\mbox{\textit{Pe}}}  
\newcommand\Ri{\mbox{\textit{Ri}}}  

\def\smallint{\begingroup\textstyle \int\endgroup}

\title{Analogy between streamers in sinking spheroids, gyrotactic plumes and chemotactic collapse}

\author{Lloyd Fung\aff{1}
  \corresp{\email{lsf27@cam.ac.uk}}
}

\affiliation{
\aff{1}DAMTP, Centre for Mathematical Sciences, Wilberforce Road, Cambridge CB3 0WA, UK}

\begin{document}
\maketitle

\input{Draft/abstract}



\input{Draft/Intro}

\input{Draft/Formulation}

\input{Draft/Analytical_Solution}

\input{Draft/Bifurcation}

\input{Draft/blowup}

\input{Draft/Discuss}



\backsection[Funding]{This work is funded by the Research Fellowship from Peterhouse, Cambridge.}

\backsection[Declaration of interests]{The authors report no conflict of interest.}


\backsection[Author ORCIDs]{L. Fung, https://orcid.org/0000-0002-1775-5093}


\appendix
\input{Draft/appex}

\bibliographystyle{jfm}
\bibliography{MyLibrary}

\end{document}

%% file: Draft/abstract.tex
\begin{abstract}
    In a dilute suspension where sinking spheroids or motile gyrotactic microorganisms are modelled as orientable and negatively buoyant particles, we have found analytical solutions to their steady distributions under any arbitrary continuous vertical shear flow. 
    The two-way coupling between their distribution and the vertical flow is nonlinear, enabling the uniform base state to bifurcate into a structure reminiscent of the streamers in settling spheroid suspensions and gyrotactic plumes.
    This bifurcation depends on a single parameter that is proportional to the average number of particles on a horizontal cross section.
    In a three-dimensional axisymmetric system, the plume structure blows up when the parameter is above a threshold. We discuss how this singularity is analogous to the chemotactic collapse of a Keller-Segel model, and the significance this analogy entails. 
  \end{abstract}

%% file: Draft/Intro.tex
\section{Introduction}
It is well-known that suspensions of sinking or motile particles can spontaneously form patterns due to instabilities in the uniform base state. 
For example, in a dilute suspension of identical sinking prolate spheroids or rods, the seminal paper by \cite{Koch1989} demonstrated the instability that gives rise to the streamer structure. 
The mechanism is as follows: a perturbation in the spatial distribution of the negatively buoyant particles creates a shear flow that attracts more particles towards regions of higher particle concentration, thereby creating a positive-feedback self-focusing mechanism and resulting in the streamer structures.
Under the assumption of a Stokes flow, \cite{Koch1989} showed that the zero wavenumbers are the most unstable, implying that there would only be a single streamer spanning the width of the container, but the experiment of \cite{Metzger2007,Metzger2007a} have shown otherwise. 
Later, \cite{Dahlkild2011}'s linear analysis showed that a finite Stokes number could regularise wavenumber, and \cite{Zhang2013} further extended the analysis nonlinearly and studied the effect of hydrodynamic diffusion.
There were also other attempts to explain the discrepancy \citep[e.g.][]{Saintillan2006}, but the wavelength selection of streamers remained an open question \citep{Guazzelli2011}.

In the meantime, monodisperse motile particle suspensions were also analysed in a similar way. \cite{Pedley1988} first demonstrated how gyrotaxis of bottom-heavy motile particles could destabilise a uniform suspension into bioconvective patterns. 
Gyrotaxis describes the tendency for bottom-heavy motile particles to swim sideways under shear due to the competing torque from the local vorticity and gravity. 
Although not recognised at the time, the mechanism of the gyrotactic instability is physically the same as the aforementioned instability and results in a similar plume structure.
Like \cite{Dahlkild2011}, \cite{Pedley1988} included the effect of finite unsteadiness due to fluid inertia. However, the finite wavelength corresponding to the most unstable mode remained larger than the experimental observation. Instead, \cite{Pedley1988} highlighted that the steady bioconvective patterns have a wavelength smaller than the initial disturbance. This phenomenon was also observed in the experiments of sinking rods \citep{Metzger2005,Metzger2007a}.

Despite the apparent similarity, gyrotactic plumes and streamers in settling spheroids/rods were historically treated as two separate topics. 
This work will report an interesting analogy between two phenomena by comparing them under the same framework, treating the buoyant and orientable particles as a continuum phase under the dilute assumption.
We will show that the gyrotactic plumes and streamers are not only physically similar but mathematically equivalent, driven by the same nonlinear particle-flow coupling that is analogous to another well-studied phenomenon known as chemotactic collapse. 
This comparative study will provide a unifying framework to compare the three phenomena, enabling an exchange of knowledge between the topics and bringing new lights to open questions on the wavelength selection of streamer structures.

%% file: Draft/Formulation.tex
\section{Formulation} \label{sec:Formulation}
\subsection{The Fokker-Planck (Smoluchowski) equations}
It has been well-established that the trajectory $\dot{\mathbf{x}}^*$ of an orientable particle (`particle' hereafter) suspended in the presence of ambient flow $\uvec^*$ can be written as
\begin{equation}
    \dot{\mathbf{x}}^* = \mathbf{u}^* + \mathbf{v}_s^*(\pvec), \label{eq:xdot_dim}
\end{equation}
where $\mathbf{v}_s^*(\pvec)$ is the slip velocity of the particle that depends on its orientation. In this work, we shall consider two kinds of particles: sinking prolate spheroids (as an approximation for rods) and spherical gyrotactic swimmers (as a model for gyrotactic motile microorganisms such as the bottom-heavy micro-algae \textit{Chlamydomonas}). For a sinking spheroid with density $\rho^* + \Delta \rho^*$, equatorial radius $a^*$ and polar length $AR a^*$ suspended in a fluid of density $\rho^*$ and viscosity $\mu^*$, the slip velocity can be written as
\begin{equation}
    \mathbf{v}_s^*(\pvec) = v_\bot^* \mathbf{e}_g+(v_\parallel^*-v_\bot^*)(\mathbf{e}_g\cdot\mathbf{p})\mathbf{p}. \label{eq:sinking_xdot}
\end{equation}
Here, 
\begin{equation}
    v_\parallel^* = \frac{2}{9}\frac{\Delta\rho^* g^* (a^*)^2 AR}{\mu^*} X (AR) \quad \mbox{and} \quad 
    v_\bot^* = \frac{2}{9}\frac{\Delta\rho^* g^* (a^*)^2 AR}{ \mu^*} Y (AR) \label{eq:sink_speeds}
\end{equation}
are the sinking speed when the particle's orientation $\pvec$, defined by the axis of revolution of the spheroid, is parallel and perpendicular to gravity $g^* \mathbf{e}_g$. 
Here, both $X (AR)$ and $Y (AR)$ are functions of the aspect ratio $AR$, the detailed formula of which can be found in appendix \ref{app:XY_form} \citep[c.f.][]{Kim1991}. 
Meanwhile, for a spherical gyrotactic swimmer, the slip velocity is 
\begin{equation}
    \mathbf{v}_s^*(\pvec) = v_\bot^* \mathbf{e}_g+v_c^* \pvec.
\end{equation}
In this work, the superscript $*$ indicates dimensional variables or parameters.

\begin{figure}
    \centering
    \includegraphics[width=0.30 \columnwidth]{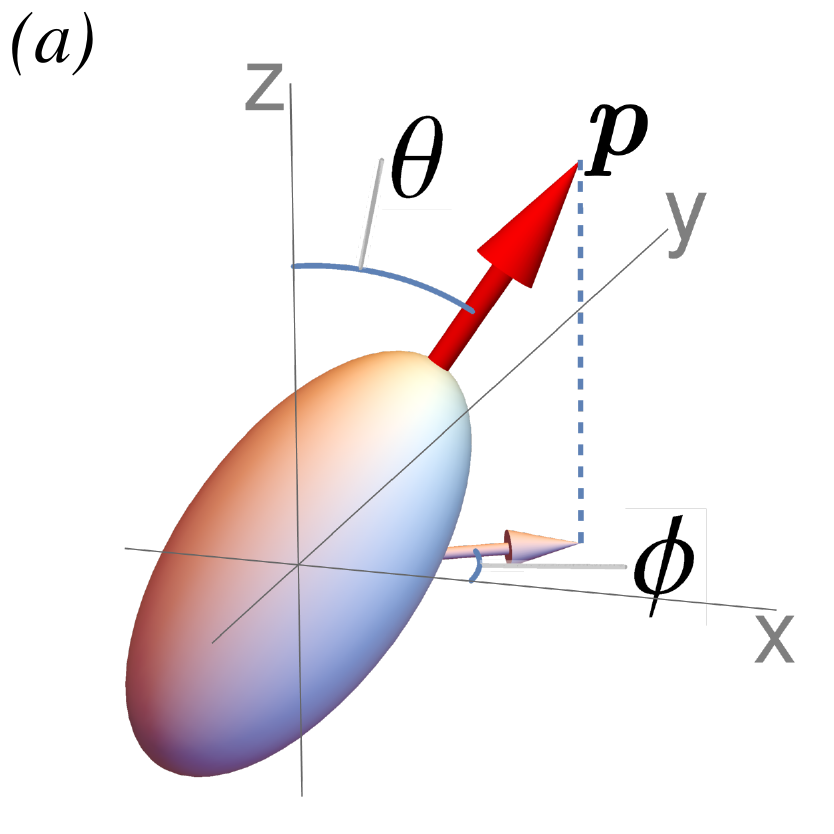}
    \includegraphics[width=0.30 \columnwidth]{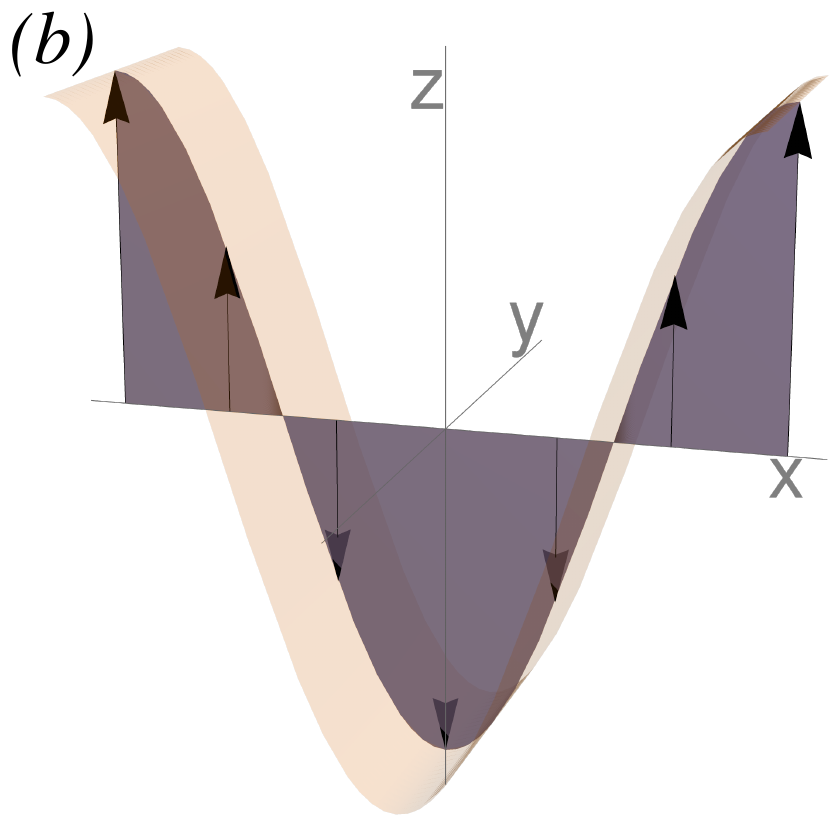} 
    \includegraphics[width=0.375 \columnwidth]{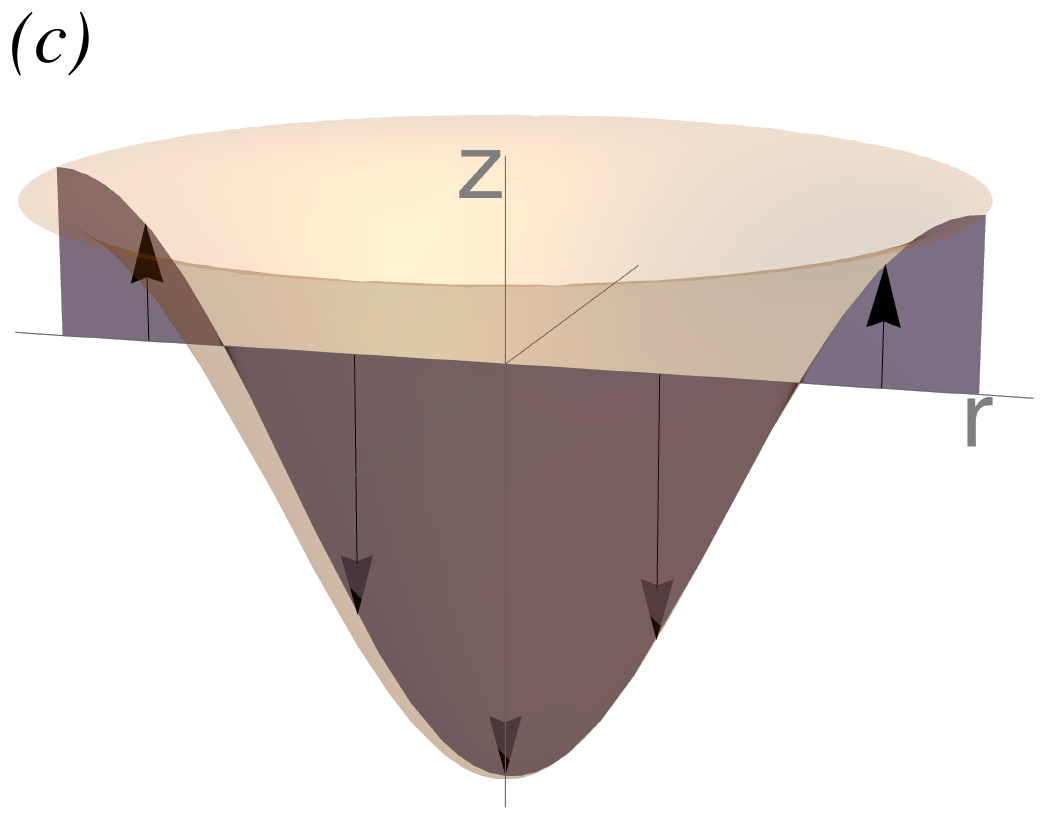}
    \caption{
        $(a)$ Diagram showing the definition of the direction $\pvec$ of a spheroid and the Euler angles representation of $\pvec$ as defined in (\ref{eq:p_def}).
        $(b)$ A typical planar vertical flow profile $u(x)$ in \S\ref{sec:bifurcation_2d} $(c)$ A typical axisymmetric vertical flow profile $u(r)$ in \S\ref{sec:blow-up}. Note that gravity is in $-z$ direction.
        \label{fig:setup}}
\end{figure}

Since the velocity $\dot{\mathbf{x}}^*$ depends on the orientation $\pvec$, the particle's angular velocity must also be resolved simultaneously. For a spheroid, the angular velocity is governed by the Jeffery orbit \citep{Bretherton1962}
\begin{equation}
    \dot{\mathbf{p}} ^*=\frac{1}{2}\mathbf{\Omega}^*\times\mathbf{p}+\alpha_0\mathbf{p}\cdot\mathbf{E}^*\cdot(\mathbf{I}-\mathbf{p}\mathbf{p}), \label{eq:pdot_dim}
\end{equation}
where $\mathbf{E}^*= \tfrac{1}{2} (\nabla^* \uvec^* + (\nabla^* \uvec^*)^T)$ and $\mathbf{\Omega}^* = \nabla^* \times \uvec^*$ are the local rate-of-strain and vorticity, and $\alpha_0=(AR^2-1)/(AR^2+1)$ is the Bretherton constant. Meanwhile, the orientational trajectory of a spherical gyrotactic particle is
\begin{equation}
    \dot{\mathbf{p}} ^*= \frac{1}{2 B^*}\left[-\mathbf{e}_g +(\mathbf{e}_g \cdot \pvec)\pvec \right] +\frac{1}{2}\mathbf{\Omega}^*\times\mathbf{p}, \label{eq:pdot_gyro_dim}
\end{equation}
where $B^*$ is the gyrotactic timescale.
The orientation $\pvec$ can also be written in terms of the Euler angles $\theta,\phi$ relative to the spatial coordinates $\mathbf{x}=(x,y,z)^T$ as shown in \ref{fig:setup}$(a)$, such that
\begin{equation} \label{eq:p_def}
    \pvec = (p_x,p_y,p_z)^T = (\sin{\theta}\cos{\phi},\sin{\theta}\sin{\phi},\cos{\theta})^T.
\end{equation}
Here, $\theta$ is the angle between $\pvec$ and the $z$-direction, which is opposite to gravity $\mathbf{e}_g$.

In a dilute and monodispersed suspension, the conservation of particles in physical and orientational space is governed by the Fokker-Planck (Smoluchowski) equation \citep{Doi1988,Saintillan2015}
\begin{equation}\label{eq:smol_dim}
\frac{\partial\Psi}{\partial t^*} + \nabla_{\textbf{x}}^*\bcdot(\dot{\mathbf{x}}^*\Psi)+\nabla_{\mathbf{p}}\bcdot(\dot{\mathbf{p}}^* \Psi)= d_r^*\nabla^2_{\mathbf{p}}\Psi ,
\end{equation}
where $\Psi(\mathbf{x}^*, \mathbf{p}, t^*)$ is the probability density function of a particle located at $\mathbf{x}^*$ with orientation $\mathbf{p}$ at time $t^*$. Here, $d_r^*$ is the rotational diffusivity, which we assume to be non-zero, homogeneous and isotropic, and represents the noise experienced by the particles. 
It models the long-range hydrodynamic disturbance from other particles or the particle's inherent thermodynamical or biological noise. Short-range interactions between particles are neglected under the dilute assumption.

The number density of particles $n(\xvec^*,t^*)$ (normalised by the average number of particles per unit volume $N^*$) can be recovered from $\Psi(\xvec^*,\pvec,t^*)$ by
\begin{equation}
    n(\xvec^*,t^*) = \int_{S_p} \Psi(\xvec^*,\pvec,t^*) d^2 \pvec,
\end{equation}
while the normalised orientational distribution can be defined as
\begin{equation}
    f(\xvec^*,\pvec,t^*) = \Psi(\xvec^*,\pvec,t^*) / n(\xvec^*,t^*), \quad \mbox{where} \quad  \int_{S_p} f(\xvec,\pvec,t^*) d^2 \pvec=1.
\end{equation}
Here, ${S_p}$ represents the spherical surface domain spanned by the orientational $\pvec$, i.e. the $\pvec$-space.
\subsection{The Navier-Stokes equations}
Meanwhile, the fluid flow $\mathbf{u^*}$ is governed by the Navier-Stokes equation
\begin{equation}
\rho^*\left( \frac{\partial \mathbf{u}^*}{\partial t^*}+\mathbf{u}^*\cdot\nabla_{\mathbf{x}}^*\mathbf{u}^*\right) = -\nabla_{\mathbf{x}}^* p^*+\mu^*\nabla_{\mathbf{x}}^{2^*}\mathbf{u}^*+\gamma^* n \mathbf{e}_g, \label{eq:NS_dim}
\end{equation}
where $p^*$ is the fluid pressure and $\gamma^* n =\Delta \rho^* g^* ({4 \pi }/{3}) (a^*)^3 AR~ N^* n$ the buoyancy force suspended particles exert on the fluid.
By writing down (\ref{eq:NS_dim}) as the equation governing the conservation of momentum in the suspension, we have implicitly assumed that the inertia of the suspended particles is negligible compared to the fluid flow $\mathbf{u}^*$.
However, the buoyancy force from the bulk of the suspended phase is significant, as shown in the last term of (\ref{eq:NS_dim}), while the higher order stress contributions from the particles are assumed relatively negligible compared buoyancy.
Together, (\ref{eq:smol_dim}) and (\ref{eq:NS_dim}) complete the set of continuum equations governing the flow in the suspension and the evolution of the particle distribution.

\subsection{Non-dimensionalisation}
To non-dimensionalise the equation, we introduce a suitable length scale $H^*$ and uses the typical particle's slip velocity $V_s^*$ to non-dimensionalise the equations.
To facilitate the analysis later, we define the typical slip velocity as $V_s^*=v_\parallel^*-v_\bot^*$ for sinking spheroids and $V_s^*=v_c^*$ for gyrotactic swimmers. 
Hence, the non-dimensionalisation gives rise to the dimensionless parameters
\begin{equation}
\Rey = \frac{\rho^* {H^*}V_s^*}{\mu^*},\,
\Ri = \frac{\gamma^* H^*}{{V_s^*}^2 \rho^*}=\frac{4 \pi}{3} \frac{\Delta \rho^*}{\rho^*} \frac{g^* (a^*)^3 AR}{{V_s^*}^2 } N^* H^* ,\,
\lambda = \frac{H^*}{2B^* V_s^*} \,
\text{and} \,\,
d_r = \frac{H^* d_r^*}{V_s^*}.
\end{equation}
Here,  $\Rey$ is the Reynolds number representing the fluid viscosity, $\Ri$ the Richardson number representing the buoyancy force from the particles (which is proportional to the number density $N^*$), and $\lambda$ the gyrotactic bias parameter.

\subsection{Applying the parallel assumption}
We consider a vertical section of width $2 H^*$ of an otherwise infinite suspension with no boundary, where a streamer (gyrotactic plume) may arise due to the instability of \cite{Koch1989} \citep{Pedley1990}. As shown in figure \ref{fig:setup}$b$, we assume the flow is always vertical (along the direction of gravity) and that there is homogeneity in the spanwise ($y$) and streamwise ($z$) direction, while periodicity is assumed in the horizontal $x$-direction. In other words, we assume $\mathbf{u} = u(x,t) \mathbf{\hat{z}}$, where the flow $u(x,t)$ only varies in the periodic domain $x \in [-1 ,1]$ and in time $t$.
Since $x$ is periodic, we further constrain $u(x,t)$ and normalise $n(x,t)$ such that 
\begin{equation} \label{eq:n-int}
    \pardx{u}= 0 \quad \mbox{at} \quad x=\pm 1, \quad \int_{-1}^1 u(x,t \,) dx=0,\quad \mbox{and} \quad  \int_{-1}^1 n(x,t)\, dx=2.
\end{equation} 
The Neumann condition implies that there is no driving pressure in $z$. Hence, (\ref{eq:NS_dim}) becomes
\begin{equation}\label{eq:NS-noG}
    \pardt{u} = \frac{1}{\text{Re}}\frac{\partial^2 u}{\partial x^2}-\text{Ri}\, (n(x,t)-1).
\end{equation}
Meanwhile, the non-dimensionalised (\ref{eq:smol_dim}) is reduced to
\begin{subequations} \label{eq:smol-parallel}
\begin{equation}
    \pardt{\Psi} +\pardx{}(K \Psi) + \mathcal{L}_{px}(x,t) \Psi= 0, 
\end{equation}
where the operation in $\pvec$-space
\begin{equation}
    \mathcal{L}_{px}(x,t) \Psi = \mathcal{L}_{px}(S(x,t)) \Psi = S(x) \mathcal{L}_{S}\Psi + \mathcal{L}_{H} \Psi  
\end{equation}
\end{subequations}
can be split into a spatially-inhomogeneous operation $S(x)\mathcal{L}_{S} \Psi$ that scales with the local shear rate $S(x,t)= \partial_x u(x,t)$, and the spatially-homogeneous operation $\mathcal{L}_{H} \Psi$. Equations (\ref{eq:smol-parallel}) are written in such a way that applying (\ref{eq:smol-parallel}) to a type of particles is now a matter of substituting the slip velocity $K$ in $x$ and the $\pvec$-space operators $\mathcal{L}_{S}$ and $\mathcal{L}_{H}$ with the corresponding particle properties. For the sinking spheroid suspension, 
\begin{subequations}\label{eq:Lop_sink}
    \begin{equation}
        K=-\cos{\theta} \sin{\theta} \cos{\phi}, \label{eq:K_sink}
    \end{equation}
\begin{eqnarray}
    \mathcal{L}_{S}\Psi  & = & \frac{1}{2} \left(
        \cot{\theta} \sin{\phi} \pardphi{\Psi} - \cos{\phi}\pardtheta{\Psi} \right) \nonumber  \\
        & + & \frac{\xi}{6} \left( 
            - 3 \cos{\phi} \sin{2 \theta} \Psi
            - \cot{\theta} \sin{\phi} \pardphi \Psi
            + \cos{2 \theta} \cos{\phi} \pardtheta \Psi
        \right) 
    \end{eqnarray}
    with $\xi=3 \alpha_0$ and
    \begin{equation}
        \mathcal{L}_{H} \Psi = - d_r \lapp \Psi  =  d_r \left(- \csc{\theta} \pardtheta{}(\sin{\theta} \pardtheta{\Psi})- \csc^2 \theta \frac{\partial^2 \Psi}{\partial \phi^2} \right),
    \end{equation}
\end{subequations}
while for gyrotactic swimmer suspension,
\begin{subequations}\label{eq:Lop_gyro}
    \begin{equation}
        K=\sin{\theta}\cos{\phi}, \label{eq:K_gyro}
    \end{equation}
\begin{equation}
    \mathcal{L}_{S}\Psi   =  \frac{1}{2} \left(
        \cot{\theta} \sin{\phi} \pardphi{\Psi} - \cos{\phi}\pardtheta{\Psi} \right) 
    \end{equation}
    and
    \begin{eqnarray}
        \mathcal{L}_{H} \Psi & = & \lambda \delp \cdot \left( [-\mathbf{e}_g + (\mathbf{e}_g \cdot \pvec)\pvec] \Psi \right) - d_r\lapp \Psi \nonumber \\
        & = & d_r \left( 2 \xi (-2 \cos{\theta} \Psi -\sin{\theta} \pardtheta{\Psi} )  -  \csc{\theta} \pardtheta{}(\sin{\theta} \pardtheta{\Psi}) -\csc^2 \theta \frac{\partial^2 \Psi}{\partial \phi^2} \right),
    \end{eqnarray}
\end{subequations}
where $\xi=\lambda/2d_r$. Here, $\xi$ is defined for each type of particle to facilitate later analyses.

%% file: Draft/Analytical_Solution.tex
\section{Analytical steady solutions to the Fokker-Planck equations} \label{sec:Analytical_Solution}
In this section, we focus on solving the steady solution of (\ref{eq:smol-parallel}) by assuming the flow has converged to a steady solution $u(x)$.
For any given arbitrary and continuous vertical flow profile $u(x)$, and thereby any arbitrary shear profile $S(x)$, the steady solution to (\ref{eq:smol-parallel}) is unique and stable as (\ref{eq:smol-parallel}) is linear and diffusive in the $\pvec$-space. Furthermore, separation of variables is possible for the given operators in (\ref{eq:Lop_sink}) and (\ref{eq:Lop_gyro}). This is because the homogeneous solution $g(\pvec)$ to the $\pvec$-space operator $\mathcal{L}_H$, i.e. $\mathcal{L}_H g(\pvec)=0$, also satisfies $\mathcal{L}_S g(\pvec)=\xi K g(\pvec)$. 
For example, in sinking spheroid suspensions, the steady separable solution has a uniform orientational distribution, i.e. uniform in the $\pvec$-space, or
\begin{equation}
    \Psi(x,\pvec, \infty)= n(x)g(\pvec) = n(x)/ 4 \pi. \label{eq:sink_analytical}
\end{equation}
Meanwhile, the steady separable solution to the gyrotactic swimmer suspension can be written as
\begin{equation}
    \Psi(x,\pvec, \infty)= n(x)g(\pvec) = n(x) \frac{2 \xi}{4 \pi \sinh{2 \xi}}\exp{(2 \xi \cos{\theta})}. \label{eq:gyro_analytical}
\end{equation}
However, it should be noted that this separation of variables is not always possible in the more general context of Fokker-Planck equations governing orientable particles. For instance, this technique cannot be applied to non-spherical gyrotactic swimmers. The critical condition that made the technique possible is when the solution to $\mathcal{L}_H g(\pvec)=0$ satisfies $\mathcal{L}_S g(\pvec)=\xi K g(\pvec)$, which both spherical gyrotactic swimmers and sinking spheroids happen to fulfil.

Now, substituting either (\ref{eq:sink_analytical}) or (\ref{eq:gyro_analytical}) into (\ref{eq:smol-parallel}) gives the same relationship between $n(x)$ and $u(x)$, that is 
\begin{equation} \label{eq:nx_steady}
    \xi S(x) n(x)= \xi u'(x) n(x)= - n'(x).
\end{equation}
In other words, sinking spheroids and gyrotactic swimmer suspension share the same steady particle distribution in $x$ for a given velocity profile $u(x)$. This is the main reason behind the analogy between gyrotactic plumes and streamers. Also, it immediately follows that
\begin{equation} \label{eq:n_analytical}
    n(x) = C \exp{\left(-\xi  u(x) \right)}
\end{equation}
where $C$ can be found by the normalisation condition (\ref{eq:n-int}). 

As mentioned above, for a given velocity profile $u(x)$, the Fokker-Planck equation (\ref{eq:smol-parallel}) on its own is a linear equation.
However, nonlinearity arises when $n(x)$ is coupled with $u(x)$ in equation (\ref{eq:NS-noG}-\ref{eq:smol-parallel}). This nonlinearity can lead to bifurcations, as we demonstrate in the next section.

%% file: Draft/Bifurcation.tex
\section{Bifurcation towards the plume/streamer structure}
\label{sec:bifurcation}
\subsection{Bifurcation in the two-dimensional case}
\label{sec:bifurcation_2d}
Armed with the explicit form of $n(x)$ at the steady state, we can numerically solve for the steady solution to the coupled equations (\ref{eq:NS-noG}-\ref{eq:smol-parallel}) governing the parallelised system.
Recall that below a certain critical Richardson number $Ri=Ri_c$, the uniform solution 
\begin{equation}
    n_0(x)=1, \quad \Psi_0=g(\pvec) n_0, \quad u_0(x)=0
\end{equation}
is a stable and steady solution to the system. However, at $Ri>Ri_c$, \cite{Koch1989} and \cite{Pedley1990} showed that the uniform basic state is prone to instability that gives rise to the streamer/plume structure. Therefore, we expect a potential bifurcation at the neutral stability point  $Ri=Ri_c$.

To demonstrate the bifurcation, we have performed numerical continuation of the steady solution at increasing $Ri$ using the numerical method described in \cite{Fung2020a}. 
\begin{figure}
    \centering
    \includegraphics[width=0.355 \columnwidth]{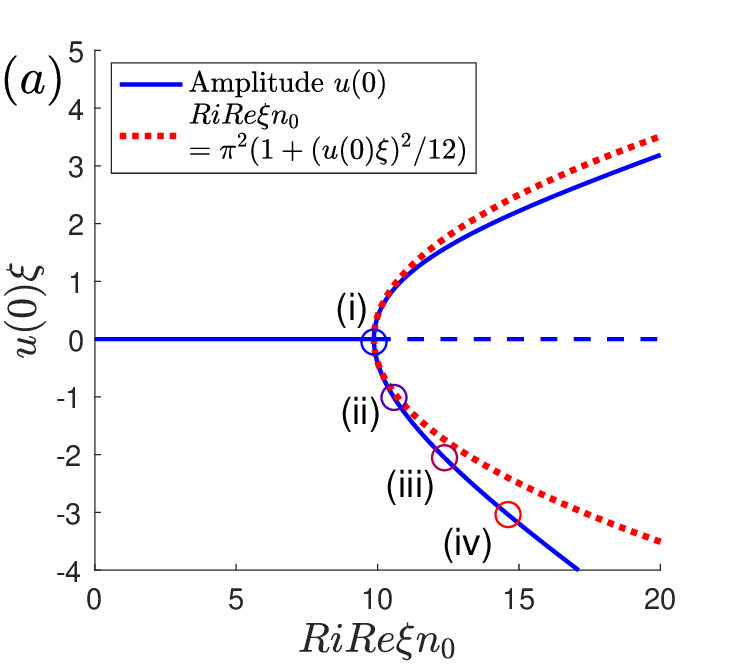}
    \includegraphics[width=0.31 \columnwidth]{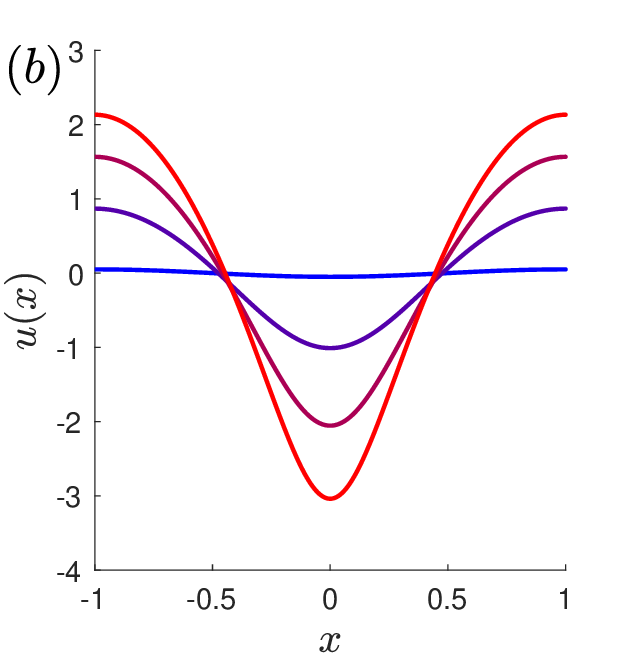} 
    \includegraphics[width=0.31 \columnwidth]{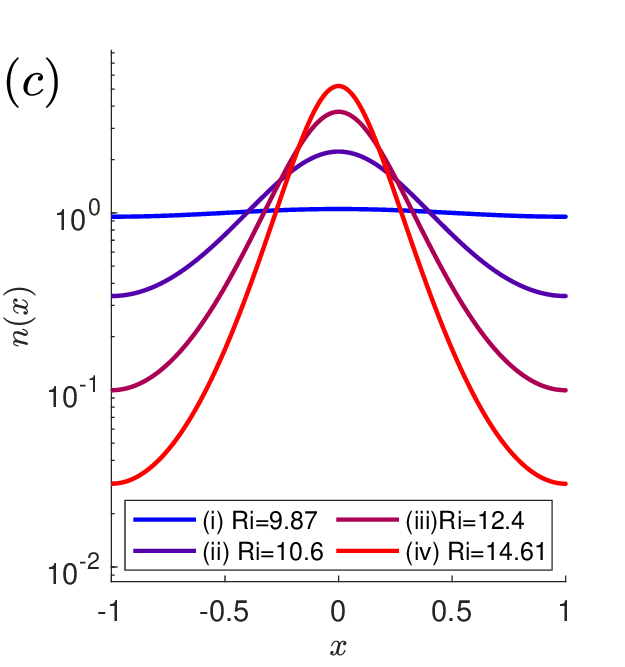}
    \caption{
        $(a)$ Bifurcation diagram on the $Ri-u(0)$ plane after rescaling. Solid line represents stable steady solution, while dashed line represents unstable steady solution. The dotted line gives the theoretical prediction from (\ref{eq:pitchfork_bi})
        $(b-c)$ Steady solutions of $(b)$ $u(x)$ and $(c)$ $n(x)$ along the continuation as marked by the circles (i-iv) in $(a)$, with $\Rey=n_0= \xi=1$.
        Note that the solutions from the upper branch in $(a)$ are equivalent to the lower branch solutions $(b,c)$ with a half-period shift in $x$. 
        \label{fig:sedi_bi}}
\end{figure}
Figures \ref{fig:sedi_bi}$(b,c)$ show some of the solutions along the lower branch, where a plume structure is clearly observed. 
Figure \ref{fig:sedi_bi}$(a)$ shows that, after rescaling $Ri$ and $u(0)$, the bifurcation collapse onto a single diagram. (Due to translational invariance in $x$, the upper branch is equivalent to the lower branch with a half-period shift in $x$. Therefore, $u(\pm 1)$ in figure \ref{fig:sedi_bi}$b$ is equivalent to $u(0)$ in the upper branch of figure \ref{fig:sedi_bi}$a$.)
Weakly nonlinear analysis in appendix \ref{app:weakly} shows that the bifurcating line on the rescaled $u(0)-\Ri$ plane can be approximated by 
\begin{equation} \label{eq:pitchfork_bi}
    {Re \xi n_0} (Ri-Ri_c) =  \frac{\pi^2}{12} \left(\xi u(0) \right)^2 \quad \mbox{with} \quad \Rey \xi n_0 Ri_c = \pi^2.
\end{equation}
In other words, there is a supercritical pitchfork bifurcation at $Ri=Ri_c=\pi^2  / \Rey \xi n_0$. 
Note that the results in figure \ref{fig:sedi_bi} would be equivalent to the numerical results of \cite{Zhang2013} if they prescribed no translational diffusion and a constant rotational diffusivity.

There are several important implications from the result. 
Firstly, the bifurcation that leads to the streamer structure depends only on a single parameter $\Rey  \xi n_0 Ri_c$. We shall delay the discussion on this parameter to \S\ref{sec:discussion}.
Secondly, we observe that the magnitude of $u(0)$ of the two possible solutions increases with $Ri$ after bifurcation but remains finite. It implies that in a two-dimensional system with infinite depth, the streamer will eventually converge to a steady structure with finite velocity and concentration at the centre. However, as we shall demonstrate below, this is not the case in a three-dimensional axisymmetric system.

%% file: Draft/blowup.tex
\subsection{Blow-up in the three-dimensional axisymmetric case} \label{sec:blow-up}
\begin{figure}
    \centering
    \includegraphics[width=0.355 \columnwidth]{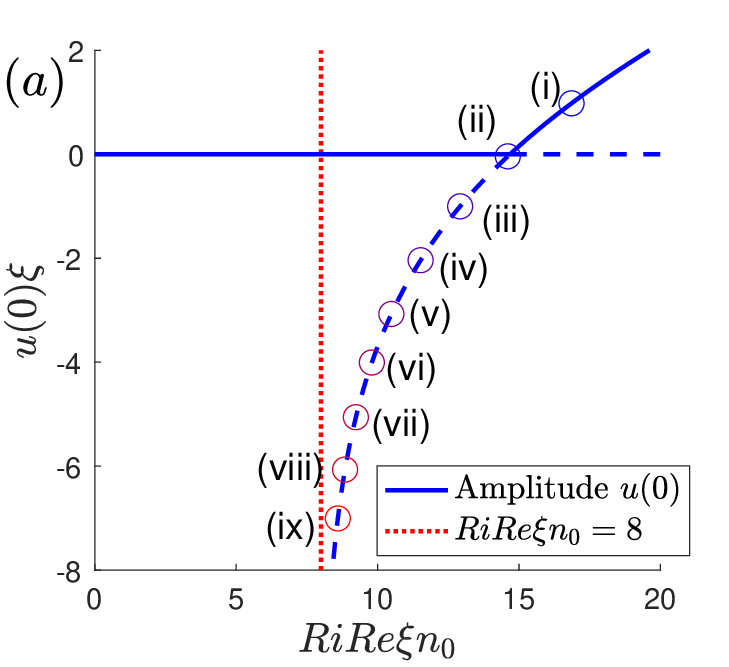}
    \includegraphics[width=0.31 \columnwidth]{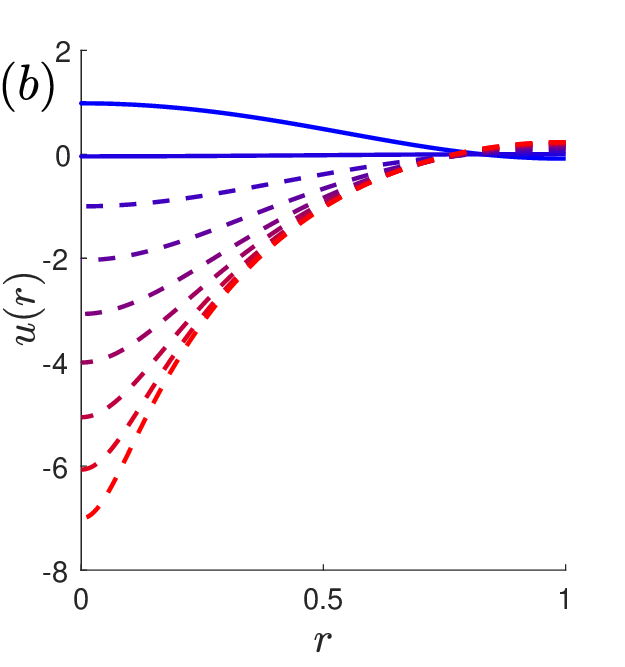} 
    \includegraphics[width=0.31 \columnwidth]{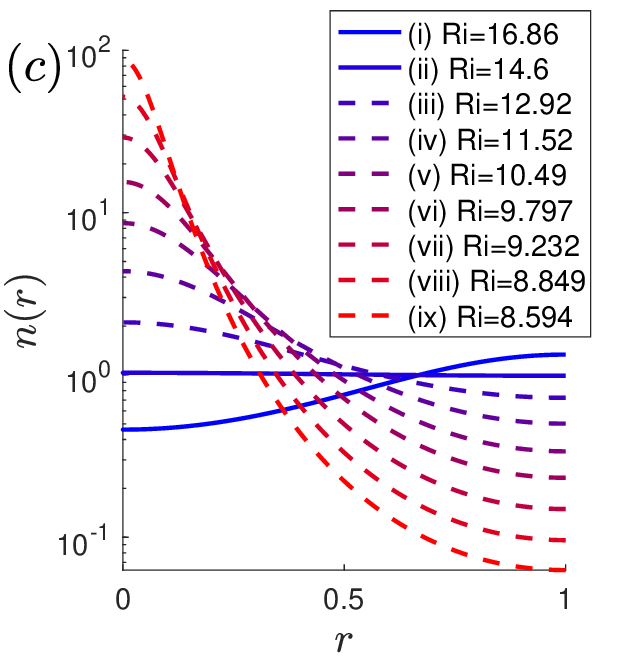}
    \caption{
        $(a)$ Bifurcation diagram on the $Ri-u(0)$ plane after rescaling. 
        $(b-c)$ Steady solutions of $(b)$ $u(r)$ and $(c)$ $n(r)$ along the continuation as marked by the circles (i-ix) in $(a)$, with $\Rey=n_0= \xi=1$.
        Here, solid line represents stable steady solution, while dashed line represents unstable steady solution.
        \label{fig:sedi_trans}}
\end{figure}

In this section, we extend the above analysis to the axisymmetric case, where we assume homogeneity in the azimuthal $\psi$ and vertical $z$ direction. Figure \ref{fig:setup}$(c)$ shows a typical axisymmetric vertical flow $u(r)$. Here, we will adopt the cylindrical coordinates $\xvec_R=(r,\psi,z)^T$ and a new set of Euler angles $\tilde{\theta}$ and $\tilde{\phi}$ as
\begin{equation}
    \pvec_R = (p_r,p_\psi,p_z)^T = (\sin{\tilde{\theta}}\cos{\tilde{\phi}},\sin{\tilde{\theta}\sin{\tilde{\phi}}},\cos{\tilde{\theta}})^T.
\end{equation}
Although the rotating $\tilde{\phi}(\psi)$ coordinate introduces a centrifugal force in the $\pvec$-space, the resulting Fokker-Planck equation is the same as (\ref{eq:smol-parallel}) with $x \mapsto r$ (see appendix \ref{app:centrifugal}). Hence, following the same procedures in \S\ref{sec:Analytical_Solution}, we have 
\begin{equation}  \label{eq:nr_steady}
    \xi u'(r) n(r)= -n'(r),
  \end{equation}
  leading to the steady solution 
\begin{equation} \label{eq:n_analytical-axis}
    n(r) = C \exp{\left(-\xi  u(r) \right)}, \quad \mbox{where} \quad \int_0^1  n(r) r dr = 1/2.
\end{equation}
Coupling it with the steady flow equation under the axisymmetric and parallel assumption,
\begin{subequations}\label{eq:NS-noG-axis}
\begin{equation} \label{eq:NS-noG-axis-a}
   0 = \frac{1}{\text{Re}}\frac{1}{r}\pardr{} (r u'(r))-\text{Ri}\, (n(r)-1),
\end{equation}
where
\begin{equation}
    u'(r)=0 \; \mbox{at} \; r=0,1 \;  \mbox{and} \; \int_0^1 u(r) r dr =0,
\end{equation}
\end{subequations}
again, we seek the steady solutions and the bifurcation numerically. 
Note that the boundary condition at $r=1$ represents a stress-free (Neumann) boundary for the flow, as we are isolating an axisymmetric plume in an infinite medium that is no longer periodic.
Also, as a result of the boundary condition, (\ref{eq:NS-noG-axis-a}) has no driving pressure gradient.
Because of the lack of translational symmetry, the bifurcation is no longer a pitchfork. Instead, it is a transcritical bifurcation, as shown in figure \ref{fig:sedi_trans}.
Linear and weakly nonlinear analysis (see appendix \ref{app:weakly_r}) show that the bifurcation points are at 
\begin{equation}
    \Rey  \xi n_0 Ri_c = \kappa^2, \quad \mbox{where} \quad J_1(\kappa)=0,
\end{equation}
where $J_n(r)$ is the $n^{th}$ Bessel function of first kind \cite[c.f.][]{Fung2020}.
Notably, the continuation from the bifurcation point in the negative $u(0)$ direction tends towards a vertical line  $Re  \xi n_0 Ri  =8$ (figure \ref{fig:sedi_trans}$a$). 
This continuation forms an unstable manifold. 
Direct dynamical simulation of the axisymmetric equivalent of (\ref{eq:NS-noG}-\ref{eq:smol-parallel}) shows that if the system is perturbed from the uniform state to beyond this manifold at $8 < Re \xi n_0 Ri< \kappa^2$, or perturbed in the negative $u(0)$ direction when $Re \xi n_0 Ri  > \kappa^2$, the number density and velocity may blow up in finite time.

%% file: Draft/Discuss.tex
\section{Discussion and concluding remark}\label{sec:discussion}
\subsection{The singularity and its connection to chemotactic collapse}
To the best knowledge of the author, this is the first demonstration of the nonlinear singularity in a suspension of sinking spheroids. 
The discovery of this singularity might provide new insights into the wavelength selection of streamer structure, as we shall discuss later. As for gyrotactic suspensions, the singularity was somewhat obscured by the development of the transport model for gyrotactic swimmers. The first analysis of gyrotactic focusing by \cite{Kessler1986} showed the singularity, but their primitive model of gyrotaxis was soon superseded by the popular FP model \citep{Pedley1990}, which suppressed the singularity. Although our recent revisit of the problem with the more accurate generalised Taylor dispersion model has rediscovered this singularity \citep{Fung2020a}, this work further supersedes our previous work by directly solving the Fokker-Planck equation, giving us confidence that the singularity is not an artefact of any transport model that approximates the equation.  Coincidentally, our result in (\ref{eq:nr_steady}) has recovered the same singular solution as \cite{Kessler1986}, although our solution is more rigorously derived.

\cite{Kessler1986} hinted at the similarity between their primitive model and Keller-Segel models for autochemotaxis. In its most simplified form \citep{Childress1981}, a Keller-Segel-type model consists of two continuum equations governing the conservation of a chemical attractant $b$ and chemotactic motile cells $a$:
\begin{equation}
    \partial_t b = \nabla^2_x b + \gamma a; \label{eq:Keller_b}
\end{equation}
\begin{equation}
    \tau \partial_t a + \bnabla_x \bcdot \left[ \chi (\bnabla_x b)a - \mu \bnabla_x a \right] = 0. \label{eq:Keller_a}
\end{equation}
The cells are producers of the attractant ($\gamma a$) in (\ref{eq:Keller_b}), but they also diffuse ($\mu \nabla^2_x a$) and drift against the chemical gradient ($\chi (\bnabla_x b) a$) at a different timescale $\tau$ in (\ref{eq:Keller_a}). Mechanistically, this is similar to how gyrotactic swimmers or sinking spheroids exert gravitational forces to accelerate the flow in the plume while being attracted to the plume due to shear (gradient in the flow velocity). Here, we shall also demonstrate their mathematical equivalence.
With $n \mapsto a$ and $u \mapsto b$, it is not difficult to see that the flow equation (\ref{eq:NS-noG}) is equivalent to (\ref{eq:Keller_b}), but the Fokker-Planck equation is more complex than (\ref{eq:Keller_a}). 
Nonetheless, the analytical solution allows us to write down (\ref{eq:nx_steady}) and (\ref{eq:nr_steady}), which is the equivalent steady solution to (\ref{eq:Keller_a}) with no-flux boundary conditions in $\mathbf{x}$. 
Therefore, the coupled Fokker-Planck and Navier-Stokes equations under the parallel and steady assumption are also a Keller-Segel-typed model.

In this light, the singularity (\S\ref{sec:blow-up}) can be interpreted as the equivalent of chemotactic collapse, a prominent feature of the Keller-Segel model. It describes the autonomous blow-up in $a$ and $b$ at a finite time and is often used to describe aggregation in biological populations, such as the aggregation of slime moulds. \cite{Childress1981} showed that chemotactic collapse is impossible in a one-dimensional system and requires a threshold number of cells in a two-dimensional system. We have shown the same in \S\ref{sec:bifurcation_2d} and \S\ref{sec:blow-up}. Physically, it is simply because, in high-dimensional systems, more particles/cells are available to amplify the positive-feedback mechanism mentioned above.

\subsection{The wavelength selection of plumes or streamers}
The singularity played an important role in the wavelength selection of chemotactic collapse. However, here it does not directly predict the wavelength of gyrotactic plumes or streamers (represented by system width $H^*$), but only constrains it with a lower bound. The analysis in \S\ref{sec:bifurcation} has shown a minimum threshold $Re \xi n_0 Ri >8$ for the uniform suspension to blow up. Expanding it back into its dimensional form for sinking spheroids
\begin{equation} \label{eq:rod_threshold}
    Re \xi n_0 Ri = a^* (H^*)^2 N^* \frac{18 \pi \alpha_0(AR)}{X(AR)-Y(AR)} = \frac{27}{2} \left(\frac{H^*}{a^*} \right)^2 c \frac{\alpha_0}{AR(X-Y)} >8
\end{equation}
shows that it is independent of the viscosity $\mu^*$, gravity $g^*$, particle density $\Delta \rho^*$ and rotational diffusivity $d_r^*$. Instead, it depends only on the aspect ratio $AR$ and the average number of particles on a horizontal cross-section of the plume $(H^*)^2 N^*$ or the volume fraction $c$. In contrast, the same level of universality cannot be said for gyrotactic swimmers, where 
\begin{equation} \label{eq:gyro_threshold}
    Re \xi n_0 Ri= \frac{1}{4 B^* d_r^*} \left(\frac{4 \pi}{3} (a^*)^3 N^* \right) \frac{(H^*)^2  \Delta \rho^* g^*}{v_c^* \mu^*}>8 . 
\end{equation}
The higher level of universality in sinking spheroid suspension is because the motility $V_s^*=v_\parallel^*-v_\bot^* \sim {\Delta\rho^* g^* (a^*)^2 AR}/{ \mu^*}$ cancels out $\gamma^*/(N^* a^*)=\Delta \rho^* g^* ({4 \pi }/{3}) (a^*)^2 AR~ $ and $\mu^*$, in contrast to the motility of swimmers.
Regardless of the particles, the physical implication of (\ref{eq:rod_threshold}-\ref{eq:gyro_threshold}) is that there exists a minimum width ($2H^*$) to the streamer/plume structure, which depends only on the background concentration $N^*$ for the given particle and fluid properties. However, this minimum width is not necessarily the wavelength of the observed pattern. For example, plugging the parametric values for \emph{C. augustae} (n\'{e}e \emph{C. nivalis}, see \citealt{Pedley1990}) into (\ref{eq:gyro_threshold}) gives a minimum plume width of $\approx 1.3mm$ at $N^*\approx 10^6 cm^{-3}$, which might be of similar order as the observed 1-3mm in bioconvection \citep{Pedley1988}. However, applying (\ref{eq:rod_threshold}) to the experiment of \cite{Metzger2007} gives a minimum streamer width of $\approx 4$ rod lengths at $0.5\%$ volume fraction, an order of magnitude smaller than the observed streamer width. 

Nonetheless, the discovery of this singularity gives a new interpretation of the wavelength selection in bioconvection and streamer structure of settling rods. 
Since local perturbations can easily trigger a chemotactic collapse, multiple plumes can arise from an initial uniform suspension, as long as there are more particles than the threshold $N^*(H^*)^2$ on the horizontal cross-section of each plume. 
As the plume blows up, the short-range or multi-particle hydrodynamic interactions neglected by the dilute assumption will likely play a significant role in regularising the singularity.
Therefore, the destiny of the streamers/plumes depends entirely upon the physics that regularises the singularity. 
For example, it might be that the regularisation in the gyrotactic plume was able to stabilise the structure, resulting in a steady bioconvective pattern \citep[see][]{Bees2020}.
However, a different regularisation in the settling rod suspension may have caused the evolving clusters in the streamers and the breakup of streamers at a later stage \citep{Metzger2007}.
It is also likely that the wavelength selection depends strongly on the hydrodynamic interaction that regularises the singularity.

Much work may follow after this analogy between the three phenomena, as it connects three separate fields under a single unifying framework and enables the transfer of knowledge between them. For example, the regularisation technique in chemotactic collapse \citep{Lankeit2020} can be used to study plumes and streamers. Previous experiments on bioconvection \citep{Bees1997} can now be compared with the sedimentation of rods \citep{Metzger2007,Metzger2007a}. Finite-depth effects on the bioconvection wavelength \citep{Hill1989,Bees1998a} can be used to re-examine the effect of the bottom wall in the sedimentation of rods \citep{Saintillan2006}. 
Although this work did not directly predict the wavelength of the streamer, it provides a new context in which short-range hydrodynamic interactions may play a role in maintaining the plumes/streamers and keeping the system from blowing up. The role short-range interaction plays in this dynamic might be akin to how volume exclusion regularises chemotactic collapse. Therefore, the analogy this work has shown may provide a new foundation for future work.

%% file: Draft/appex.tex
\section{Formula for the sedimentation speed of spheroids} \label{app:XY_form}
By symmetry of the particle, the sinking speed of a spheroid is characterised by two resistance function $X^A$ and $Y^A$. When the external force $F^*$ applied is parallel to the axis of symmetry of the particle,
\begin{equation}
    F^* = 6 \pi \mu^* a^* AR \; X^A v_\parallel^* .
\end{equation}
When the external force $F^*$ applied is perpendicular to the axis of symmetry of the particle,
\begin{equation}
    F^* = 6 \pi \mu^* a^* AR \; Y^A v_\bot^* .
\end{equation}
Taken from \citet[][table 3.4]{Kim1991}, the resistance functions $X^A$ and $Y^A$ relating the force due to translation along and perpendicular to the axis of symmetry of the particles are
\begin{equation}
    X^A=\frac{8}{3} e^3 \left[ -2e + (1+e^2) L \right]^{-1} \: \mbox{and} \: Y^A=\frac{16}{3} e^3 \left[ 2e + (3e^2-1) L \right]^{-1}, 
\end{equation}
where
\begin{equation}
    e=\sqrt{1-\frac{1}{AR^2}}
\end{equation}
is the eccentricity of the spheroid and 
\begin{equation}
    L=\ln{(\frac{1+e}{1-e})}.
\end{equation}
Therefore, balancing the resistive force with gravity results in $X=(AR~X^A)^{-1}$ and $Y=(AR~Y^A)^{-1}$. An alternative formula can also be found in \cite{Cabrera2022}.

\section{Linear and weakly nonlinear analysis in planar coordinates} \label{app:weakly}
Here, we demonstrate that the bifurcation in \S\ref{sec:bifurcation} is a supercritical pitchfork bifurcation. 
We define $Ri=Ri_c + \epsilon^2 \Delta Ri$, slow time $T=\epsilon^2 t$, and expand $u$ and $\Psi$ as
\begin{eqnarray}
    u &=& 0 + \epsilon u_1 + \epsilon^2 u_2 + \epsilon^3 u_3 + ... \\
    \Psi &=& g(\pvec)n_0 + \epsilon \Psi_1 + \epsilon^2 \Psi_2 + \epsilon^3 \Psi_3 + ...
\end{eqnarray}
where we also define $n_i= \smallint_{S_p} \Psi_i d^2 \pvec$.
Substituting the above into (\ref{eq:NS-noG}) and (\ref{eq:smol-parallel}) and collecting the terms at each order, we have, at the first order $\mathcal{O}(\epsilon)$,
\begin{subequations}\label{eq:lin}
    \begin{eqnarray}
        \partial_t u_1 &=& \Rey^{-1} \mathcal{D}^2 u_1 - Ri \, n_1, \\
        \partial_t \Psi_1 &=& - \xi K g(\pvec)n_0 \mathcal{D} u_1 - K \mathcal{D}\Psi_1 - \mathcal{L}_H \Psi_1,
    \end{eqnarray}
\end{subequations}
where $\mathcal{D}=\partial / \partial x$.
At $Ri=Ri_c$, the neutral stability $\partial_t u_1 = \partial_t \Psi_1 = 0$ leads to 
\begin{equation}
    \mathcal{D}^2 u_1 + Re Ri \xi n_0 \; u_1 = 0. 
\end{equation}
Solving the equation with the boundary condition gives the value of $Ri_c$ and stability mode
\begin{equation}
    u_1 =A \cos {(\pi x)} \quad \mbox{and} \quad
    \Psi_1 = -A g(\pvec) { \xi n_0} \cos{(\pi x)},
\end{equation}
where $A=A(T)$ is the amplitude of the linear mode growing in the slow timescale $T$. 
The next order is degenerate due to translational invariance, so the bifurcation is demonstrated at $\mathcal{O}(\epsilon^3)$, where
\begin{subequations}\label{eq:3rd_order}
    \begin{eqnarray}
        \partial_T {u_1} + \Delta Ri_c n_1 &=& {\Rey}^{-1} \mathcal{D}^2 u_3- Ri \, n_3 \\
        \partial_T {\Psi_1}  + (\mathcal{D} u_1) \mathcal{L}_S \Psi_2 + (\mathcal{D} u_2) \mathcal{L}_S \Psi_1 &=& -  \xi K \Psi_0 \mathcal{D} u_3 -  K \mathcal{D}\Psi_3 - \mathcal{L}_H \Psi_3 .
    \end{eqnarray}
\end{subequations}
By Fredholm alternative and $A'(T)=0$, we can show that 
\begin{equation}
    -\xi n_0 \Delta Ri A + \frac{\pi^2  \xi^2}{12 \Rey}A^3 =0 \quad \mbox{or} \quad  \Delta Ri = \frac{\pi^2 \xi}{12 n_0 \Rey}  A^2 = \frac{Ri_c}{12}(\xi A)^2.
\end{equation}
In other words, there is a supercritical pitchfork bifurcation at $Ri=Ri_c$.

\section{The Fokker-Planck equation in cylindrical coordinates}\label{app:centrifugal}
In this section, we demonstrate how to convert the Fokker-Planck equation
\begin{equation}\label{eq:smol}
    \frac{\partial\Psi}{\partial t} + \nabla_{\textbf{x}}\bcdot(\dot{\mathbf{x}}\Psi)+\nabla_{\mathbf{p}}\bcdot(\dot{\mathbf{p}} \Psi)= d_r\nabla^2_{\mathbf{p}}\Psi ,
\end{equation}
in Cartesian coordinates, which governs $\Psi=\Psi(t,\xvec,\pvec)=\Psi(t,x,y,z,\theta,\phi)$, into an equivalent equation governing $\Psi=\tilde{\Psi}(t,\xvec_R,\pvec_R)=\tilde\Psi(t,r,\psi,z,\tilde{\theta},\tilde{\phi})$ in cylindrical coordinates.
First, we note that
\begin{equation}
    \phi=\psi+\tilde{\phi},
\end{equation}
and therefore
\begin{equation}
    \frac{1}{r} \left. \frac{\partial \Psi}{\partial \psi} \right|_{\phi}
    = \frac{1}{r} \left( \left.\frac{\partial \tilde\Psi}{\partial \psi} \right|_{\tilde\phi} \left. \frac{\partial \psi}{\partial \psi} \right|_{\phi} + \left. \frac{\partial \tilde\Psi}{\partial \tilde\phi} \right|_{\psi} \left.\frac{\partial \tilde{\phi}}{\partial \psi} \right|_{\phi} \right)
    = \frac{1}{r} \left( \left.\frac{\partial \tilde\Psi}{\partial \psi} \right|_{\tilde\phi} - \left. \frac{\partial \tilde\Psi}{\partial \tilde\phi} \right|_{\psi} \right).
\end{equation}
Substituting the above while converting $\xvec$-space divergence into cylindrical coordinates gives
\begin{equation} \label{eq:spatial-div-convert}
    \bnabla_x \bcdot \left[\dot\xvec \Psi \right]_\pvec 
    = \dot\xvec \bcdot \bnabla_\xvec \Psi |_\pvec 
    = \dot{\xvec}_R \bcdot  \tilde\Psi |_{\pvec_R} - \frac{\dot{x}_\psi}{r} \left. \frac{\partial \tilde\Psi }{\partial \tilde\phi } \right|_{\psi}
    =\bnabla_{\xvec_R} \bcdot \left[ \dot{\xvec}_R \tilde{\Psi} \right]_{\pvec_R} 
    -\frac{\dot{x}_r \tilde\Psi}{r} 
    - \frac{\dot{x}_\psi}{r} \left. \frac{\partial \tilde\Psi }{\partial \tilde\phi } \right|_{\psi},
\end{equation}
where in the last step, the term $\dot{x}_r \tilde{\Psi}/r$ arise from $(\bnabla_{\xvec_R} \bcdot \dot{\xvec}_R)\tilde{\Psi}$. Here,  $\dot{x}_r=-p_z p_r$ and $\dot{x}_\psi=-p_z p_\psi$ for sinking spheroids and $\dot{x}_r = p_r$ and $\dot{x}_\psi = p_\psi$ for gyrotactic swimmers.

Meanwhile, the $\pvec$-space (angular) velocity can be decomposed into 
\begin{equation}
    \dot\pvec = \dot{\pvec}_R + \frac{\dot{x}_\psi}{r} \hat{\mathbf{z}} \times \pvec_R,
\end{equation}
where the last term represents the centrifugal force arising from the angular velocity $\dot{x}_\psi$ of the rotating $\pvec_R$-space. 
Meanwhile, the operator $\bnabla_\pvec = \bnabla_{\pvec_R} $ remains the same after the change in coordinates as it was operating at constant $\xvec=\xvec_R$. Hence, the laplacian
\begin{equation} \label{eq:lapp-r}
    \nabla^2_{{\pvec}} \Psi = \nabla^2_{\pvec_R} \tilde{\Psi}
\end{equation}
also remains unchange. 
Substituting above while converting $\pvec$-space divergence from $\pvec$-space to the rotating $\pvec_R$-space gives
\begin{equation} \label{eq:pspace-div-convert}
    \bnabla_\pvec \bcdot [\dot\pvec \Psi] 
    = \bnabla_{\pvec_R} \bcdot [\dot{\pvec} \tilde{\Psi}] 
    = \bnabla_{\pvec_R} \bcdot [\dot{\pvec}_R + \frac{\dot{x}_\psi}{r} \hat{\mathbf{z}} \times \pvec_R \tilde{\Psi}]
    = \bnabla_{\pvec_R} \bcdot [\dot{\pvec}_R \tilde{\Psi}]    +\frac{\dot{x}_r \tilde\Psi}{r} 
    + \frac{\dot{x}_\psi}{r} \frac{\partial \tilde\Psi }{\partial \tilde\phi } ,
\end{equation}
The last two terms arising from the centrifugal force in (\ref{eq:pspace-div-convert}) will therefore cancel out with the last two term in (\ref{eq:spatial-div-convert}) when we substitute (\ref{eq:pspace-div-convert},\ref{eq:lapp-r}-\ref{eq:spatial-div-convert}) into (\ref{eq:smol}), resulting in
\begin{equation}\label{eq:smol_cylindrical_full}
    \frac{\partial \tilde\Psi}{\partial t} + \bnabla_{\xvec_R}\bcdot(\dot{\xvec}_R\tilde{\Psi})+\bnabla_{\pvec_R}\bcdot(\dot{\pvec}_R \tilde\Psi)= d_r\nabla^2_{\pvec_R} \tilde\Psi ,
\end{equation}
which is the Fokker-Planck equation written in cylindrical coordinates. The derivated equation is also consistent with equations (2.13-2.17) in \cite{Jiang2020}.

In practice, (\ref{eq:smol_cylindrical_full}) is rarely used directly, as it involves duplicated terms and unintuitive expansion of $\dot{\pvec}_R$. Instead, we use the intermediate result
\begin{equation}\label{eq:smol_cylindrical}
    \frac{\partial \tilde\Psi}{\partial t} + 
    \dot{\xvec}_R \bcdot  \tilde\Psi |_{\pvec_R} 
    - \frac{1}{r} \left. \frac{\partial \tilde\Psi }{\partial \tilde\phi } \right|_{\psi} 
    +\bnabla_{\pvec_R} \bcdot \left( \dot{\pvec} \tilde\Psi \right) = d_r\nabla^2_{\pvec_R} \tilde\Psi,
\end{equation}
where $\dot{\pvec}$ can be represented in terms of $(\tilde{\theta}$,$\tilde{\phi})$ and the gradient of $\mathbf{u}$ written in cylindrical coordinate. The main advantage of (\ref{eq:smol_cylindrical}) over (\ref{eq:smol_cylindrical_full}) is that the duplicate term $\dot{x}_r \tilde\Psi/r$ is already cancelled out, and that formulas for $\dot{\pvec}$ in terms of $(\tilde{\theta}$,$\tilde{\phi})$ are more readily available.

Now, in the axisymmetric case where $\tilde{\Psi}=\tilde{\Psi}(r,\pvec_R,t)$ and $\mathbf{u}=u(r,t)\mathbf{\hat{z}}$, and the Fokker-Planck equation becomes
\begin{equation}
    \pardt{\tilde\Psi}+ \dot{x}_r  \pardr{\tilde\Psi}+ \bnabla_{\pvec_R} \bcdot \left( \dot{\pvec} \tilde\Psi \right) = d_r \nabla^2_{\pvec_R} \tilde\Psi. \label{eq:smol_rsimp}
\end{equation}
In particular, for the examples considered in this work, (\ref{eq:smol_cylindrical}) can be written as
\begin{equation} \label{eq:smol-parallel-r}
    \pardt{\tilde\Psi} +\tilde{K} \pardr{\tilde\Psi} + \tilde{\mathcal{L}}_{pr}(r,t) \tilde\Psi= 0, 
\end{equation}
where
\begin{subequations} \label{eq:Lop_sink-r}
    \begin{equation}
        \tilde{K} =-\cos{\tilde\theta} \sin{\tilde\theta} \cos{\tilde\phi}, \label{eq:K_sink-r}
    \end{equation}
    \begin{eqnarray}
        \tilde{\mathcal{L}}_{pr}(r,t) \tilde\Psi & = & \left( \frac{\partial u}{\partial r} \right) \left[ \frac{1}{2} \left(
            \cot{\tilde\theta} \sin{\tilde\phi} \frac{\partial \tilde\Psi}{\partial \tilde\phi} - \cos{\tilde\phi} \frac{\partial \tilde\Psi}{\partial \tilde\theta} \right) \right. \nonumber  \\
            & + & \left. \frac{\xi}{6} \left( 
                - 3 \cos{\tilde\phi} \sin{2 \tilde\theta} \tilde\Psi
                - \cot{\tilde\theta} \sin{\tilde\phi} \frac{\partial \tilde\Psi}{\partial \tilde\phi}
                + \cos{2 \tilde\theta} \cos{\tilde\phi} \frac{\partial \tilde\Psi}{\partial \tilde\theta}
            \right) \right] \nonumber   \\
            & - & d_r \nabla_{\pvec_R}^2 \tilde\Psi
        \end{eqnarray}
\end{subequations}
with $\xi=3 \alpha_0$ for the sinking spheroid suspension, and
\begin{subequations}\label{eq:Lop_gyro-r}
    \begin{equation}
        \tilde{K}=\sin{\tilde\theta}\cos{\tilde\phi}, \label{eq:K_gyro-r}
    \end{equation}
    \begin{eqnarray}
    \tilde{\mathcal{L}}_{pr}(r,t) \tilde\Psi 
    & = & \frac{1}{2}  \frac{\partial u}{\partial r} \left(
        \cot{\tilde\theta} \sin{\tilde\phi} \frac{\partial \tilde\Psi}{\partial \tilde\phi} - \cos{\tilde\phi} \frac{\partial \tilde\Psi}{\partial \tilde\theta}\right) \\
    & & + d_r \left( 2 \xi (-2 \cos{\tilde\theta} \tilde\Psi -\sin{\tilde\theta} \frac{\partial \tilde\Psi}{\partial \tilde\theta} )  -  \nabla^2_{\pvec_R} \tilde\Psi \ \right)
    \end{eqnarray}
\end{subequations}
with $\xi=\lambda/2d_r$ for gyrotactic swimmer suspension.

The above equations are effectively the same as (\ref{eq:smol-parallel}-\ref{eq:Lop_gyro}) if we take $r \mapsto x$, $\partial_r \mapsto \partial_x$, $\tilde{\theta} \mapsto \theta$, $\tilde{\phi} \mapsto \phi$ and $\tilde\Psi \mapsto \Psi$. 
Hence, one can follow the same procedure from (\ref{eq:smol-parallel}) to (\ref{eq:nx_steady}-\ref{eq:n_analytical}) to get (\ref{eq:nr_steady}-\ref{eq:n_analytical-axis}).

\section{Linear and weakly nonlinear analysis in cylindrical coordinates}\label{app:weakly_r}
In this section, we shall demonstrate, through weakly nonlinear analysis, that the bifurcation is a supercritical pitchfork bifurcation. 
We define $Ri=Ri_c + \epsilon \Delta Ri$, slow time $T=\epsilon t$, and expand $u$ and $\Psi$ as
\begin{eqnarray}
    u &=& 0 + \epsilon u_1 + \epsilon^2 u_2 + \epsilon^3 u_3 + ... \\
    \Psi &=& g(\pvec)n_0 + \epsilon \Psi_1 + \epsilon^2 \Psi_2 + \epsilon^3 \Psi_3 + ...
\end{eqnarray}
which leads to
\begin{equation}
    n = 1 + \epsilon n_1 + \epsilon^2 n_2 + \epsilon^3 n_3 + ...
\end{equation}
where we define $n_i= \int_{S_p} \Psi_i d^2 \pvec$.
Substituting the above into (\ref{eq:NS-noG}) and (\ref{eq:smol-parallel}) of the paper and collecting the terms at each order, we have, at the first order $\mathcal{O}(\epsilon)$,
\begin{subequations}\label{eq:lin_r}
    \begin{eqnarray}
        \pardt{u_1} &=& \frac{1}{\Rey}\mathcal{D}^2 u_1 - Ri \, n_1, \\
        \pardt{\Psi_1} &=& - \xi K g(\pvec)n_0  \mathcal{D} u_1 - K \mathcal{D}\Psi_1 -\mathcal{L}_H \Psi_1,
    \end{eqnarray}
\end{subequations}
where $\mathcal{D} = \partial / \partial r$ and $\mathcal{D}^2 = (1/r)(\partial / \partial r)(r \partial / \partial r)$. 
At $Ri=Ri_c$, $\partial_t u_1 = \partial_u \Psi_1 = 0$, which leads to 
\begin{equation}
    \mathcal{D}^2 u_1 + {Re Ri \xi n_0} u_1 = \mathcal{D}^2 u_1 + \kappa^2 u_1 = 0, \quad \mbox{where} \quad \kappa^2= {Re Ri_c \xi n_0} .
\end{equation}
Solving the equation with the boundary conditions gives the stability mode
\begin{equation}
    u_1 =A J_0(\kappa r) \quad \mbox{and} \quad
    \Psi_1 = -A g(\pvec) \xi n_0 J_0(\kappa r),
\end{equation}
where $A=u_1(0)=A(T)$ is the amplitude of the linear mode growing in the slow timescale $T$, and $J_m(r)$ the $m^{th}$ Bessel functinoo fthe first kind, and $\kappa$ the first zero of $J_1(r)$. 
At the next order $\mathcal{O}(\epsilon^2)$, we have
\begin{subequations}
    \begin{eqnarray}
        \pardT{u_1} + \Delta Ri_c n_1 &=& - \pardt{u_2} + \frac{1}{\Rey}\mathcal{D}^2 u_2 - Ri \, n_2 \\
        \pardT{\Psi_1} + (\mathcal{D} u_1) \mathcal{L}_S \Psi_1 &=&  -\pardt{\Psi_2}  - \xi K \Psi_0 \mathcal{D} u_2 - \Pe_s K \mathcal{D}\Psi_2 -\mathcal{L}_H \Psi_2. 
    \end{eqnarray}
\end{subequations}
Note that $\mathcal{L}_S \Psi_1 = \xi K \Psi_1$ (see \S\ref{sec:Analytical_Solution}). 
When the solution reaches a steady saturation, $A'(T)=0$ and $\partial_t u_2 = \partial_u \Psi_2 = 0$. Hence,
\begin{subequations}\label{eq:2nd_order}
    \begin{eqnarray}
         \Delta Ri_c n_1 &=&  \frac{1}{\Rey}\mathcal{D}^2 u_2 - Ri \, n_2 \\
         (\mathcal{D} u_1) \mathcal{L}_S \Psi_1 &=&  - \xi K \Psi_0 \mathcal{D} u_2 - \Pe_s K \mathcal{D}\Psi_2 -\mathcal{L}_H \Psi_2. 
    \end{eqnarray}
\end{subequations}
Here, Fredholm alternative is automatically satisfied in the $\pvec$-space, while  the Fredholm alternative in the $r$-space requires
\begin{equation}
    \int_{0}^1 v \cdot \left( \Delta Ri \, n_1 \right) r dr + \int_{0}^1  \Phi \cdot \left((\mathcal{D} u_1) \xi K \Psi_1 \right)  r dr = 0,
\end{equation}
where $v(r,\pvec)$ and $\Phi(r,\pvec)$ are the solutions to the adjoint of the right-hand side of (\ref{eq:lin_r}).
Therefore, we can show that
\begin{equation}
    - C{\xi n_0}\Delta Ri A + E \frac{\kappa^2}{\Rey n_0}A^2 =0
\end{equation}
or
\begin{equation}
    \Delta Ri = \frac{E}{C} Ri_c \xi A,
\end{equation}
where $C$ and $E$ are defined as
\begin{equation}
    C = \int_0^1 [J_0(\kappa r)]^2 r dr = \frac{1}{2} [J_1(\kappa)]^2
\end{equation}
and
\begin{equation}
    E = \int_0^1 J_0(\kappa r) J_1^2(\kappa r) r dr.
\end{equation}
In other words, there is a transcritical bifurcation at $Ri=Ri_c$.